\documentclass[sigconf]{acmart}

\usepackage{tabularx}

\mathchardef\mhyphen="2D

\AtBeginDocument{%
  \providecommand\BibTeX{{%
    \normalfont B\kern-0.5em{\scshape i\kern-0.25em b}\kern-0.8em\TeX}}}

\setcopyright{acmcopyright}
\copyrightyear{2021}
\acmYear{2021}
\acmDOI{10.1145/1122445.1122456}

\acmConference[KDD 2021 (to appear)]{27th ACM SIGKDD Conference on Knowledge Discovery and Data Mining}{14--18 August, 2021}{Virtual}
\acmBooktitle{Proceedings of the 27th ACM SIGKDD Conference on Knowledge Discovery and Data Mining,
  14--18 August, 2021, Virtual}
\acmPrice{15.00}
\acmISBN{978-1-4503-XXXX-X/18/06}

\begin{document}

\title{Mondegreen: A Post-Processing Solution to Speech Recognition Error Correction for Voice Search Queries}




\author{Sukhdeep S. Sodhi, Ellie Ka-In Chio, Ambarish Jash, Santiago Ontañón}
\authornote{S. Sodhi, E. Chio, A. Jash and S. Ontañón had equal contribution.}
\author{Ajit Apte, Ankit Kumar, Ayooluwakunmi Jeje, Dima Kuzmin, Harry Fung, Heng-Tze Cheng, Jon Effrat, Tarush Bali, Nitin Jindal, Pei Cao, Sarvjeet Singh, Senqiang Zhou, Tameen Khan, Amol Wankhede, Moustafa Alzantot, Allen Wu, Tushar Chandra}

\email{[sodhi,echio,ajash,santiontanon]@google.com}
\affiliation{%
  \institution{Google Research}
   \country{USA}
}








\renewcommand{\shortauthors}{Sodhi, Chio, Jash, Ontañón et al.}


\begin{abstract}
As more and more online search queries come from voice, automatic speech recognition becomes a key component to deliver relevant search results. Errors introduced by {\em automatic speech recognition} (ASR) lead to irrelevant search results returned to the user, thus causing user dissatisfaction.
In this paper, we introduce an approach, {\em Mondegreen}, to correct voice queries in text space without depending on audio signals, which may not always be available due to system constraints or privacy or bandwidth (for example, some ASR systems run on-device) considerations. 
We focus on voice queries transcribed via several proprietary commercial ASR systems. These queries come from users making internet, or online service search queries.
We first present an analysis showing how different the language distribution coming from user voice queries is from that in traditional text corpora used to train off-the-shelf ASR systems.
We then demonstrate that Mondegreen can achieve significant improvements in increased user interaction by correcting user voice queries in one of the largest search systems in Google.
Finally, we see Mondegreen as complementing existing highly-optimized production ASR systems, which may not be frequently retrained and thus lag behind due to vocabulary drifts.
\end{abstract}





\maketitle

\begin{figure}[t!]
	\includegraphics[width=0.9\columnwidth]{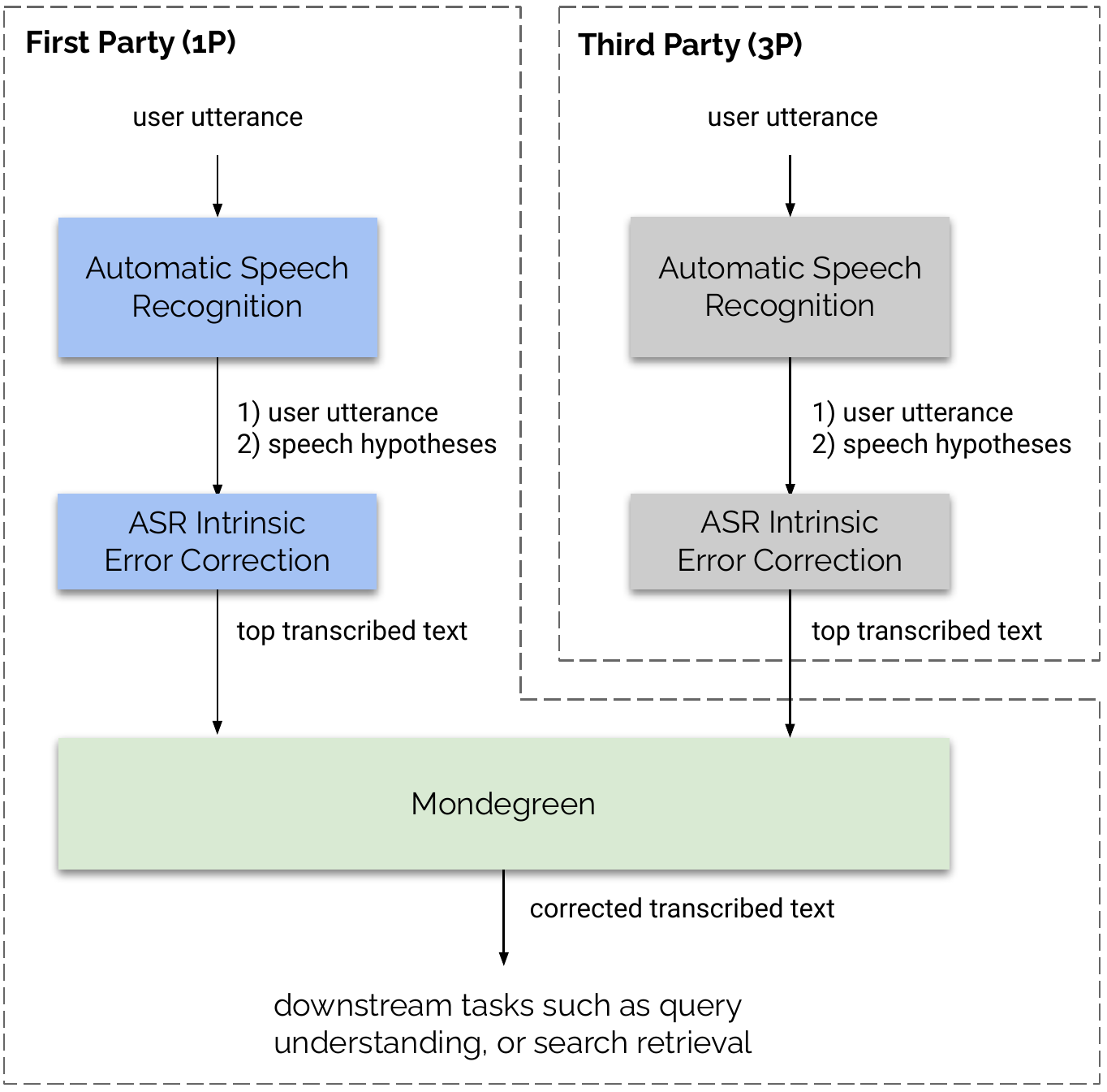}
	\centering
	\caption{A system diagram showing how Mondegreen can be used with both first-party (1P) and third-party (3P) automatic speech recognition (ASR) systems. User utterance is first transcribed by the 1P or 3P ASR, and then error correction follows. Lastly, the most probable transcription {\em without the user utterance} is passed downstream to Mondegreen, which then outputs the corrected transcribed text.} 
	\label{fig:diagram}
\end{figure}

\section{Introduction}



In {\em voice user interfaces} (VUI)~\citep{myers2018patterns},
errors introduced by {\em automatic speech recognition} (ASR) are propagated to downstream tasks such as query understanding. As VUI are becoming more widespread, ASR error detection and correction has become a key open challenge to serve voice queries faithfully. Transcription errors introduced by ASR increase the percentage of users being served irrelevant search results, which is not a problem in text-based interfaces.

In the past decade, there has been a significant reduction in ASR {\em word error rate}. For instance, in the ``testother'' benchmark of the LibreSpeech dataset, the word error rate is reduced from 12.51\%~\citep{ko2015audio} in 2015 to about 4.1\% in 2020~\citep{han2020contextnet}. 

However, as voice queries in VUIs exhibit significantly different language patterns from those that the ASR systems are usually trained on, even with these significant improvements, word error rate remains high. This leads to ``query abandonment'', which happens when users do not end up engaging with any of the search results resulting from the query in real world scenarios where the data distribution is different from standard benchmarks.

To exacerbate the aforementioned issue further, due to system constraints, bandwidth considerations (e.g., on-device speech recognition) or privacy considerations, user utterances may not be available for correcting transcription errors. When transcribing user utterances using a {\em third party} (3P) ASR, only the transcribed text will be passed to the downstream tasks.
Even in the case of using a {\em first party} (1P) ASR,
user utterances may not be available for reasons mentioned earlier. As illustrated in Figure \ref{fig:diagram}, only transcribed text is passed down to downstream systems. Different ASR systems might implement different error correction, candidate ranking or other system-specific steps (which we labeled ``ASR Intrinsic Error Correction'') which do take into account the utterance (audio signal), and produce a final single text transcription.

This paper presents {\em Mondegreen}, a query error correction system that has been deployed to correct user voice queries without depending on audio signals in one of the largest search systems from Google.
Mondegreen is designed to work not just with 1P ASR, but 3P ASRs as well. As a matter of fact, as we will show in the experimental results below, we observe most of the gains come from correcting voice queries coming from 3P ASR systems. 

Mondegreen is designed to complement, not replace, existing highly optimized ASR systems by providing very high precision corrections.
At its core, Mondegreen is a statistical approach based on estimating the probability that certain query corrections are more likely to result in successful search results than the original transcription coming from the ASR system. In order to calculate these statistics, Mondegreen is trained on a large corpus of queries collected from one of the search systems at Google.

Specifically, our results show the following:
\begin{itemize}
    \item We highlight that standard corpora to train ASR systems have very different language distributions than the language observed in voice user interfaces. The implication is that even if ASR systems are getting increasingly accurate, voice user queries are very different statistically from standard text, potentially leading to higher ASR error rate.
    \item Mondegreen's statistical approach can complement existing ASR systems without requiring to retrain or fine-tune them to improve their performance. Specifically, we show that Mondegreen can increase measures such as ``user interaction'' (percentage of queries that result in the user engaging with one of the search results for a large amount of time) by 9.49\%, and reduce the number of queries for which the user needs to issue a refinement (an alternative query hoping to get better results) by 7.09\%. 
    \item Most of the gains obtained by Mondegreen tend to come from correcting queries coming from third-party ASR systems (notice these systems are commercial-grade ASR systems supported by large companies). We hypothesize that this is due to these systems being trained in corpora with different language distributions than voice queries.
    \item We evaluate Mondegreen in different English locales (US, UK, India) and show that they have enough commonalities so that training only on US data, those commonalities are enough as for seeing gains in the other locales (training on a larger set of locales is part of our ongoing work).
\end{itemize}


The remainder of this paper is structured as follows. Section \ref{sec:related} presents related work on ASR error correction. Section \ref{sec:task} presents the task and a comparison of the language distributions on standard ASR text corpora versus voice user queries. Section \ref{sec:mondegreen} presents Mondegreen. Sections \ref{sec:experiments-neural} and \ref{sec:experiments} present our experimental results.

\section{Related Work}\label{sec:related}

Although the accuracy of speech recognition systems has increased dramatically in the past few years~\citep{mohamed2019transformers}, ASR systems still have a non-negligible error rate in many circumstances, such as when there is background noise~\citep{zhang2018deep}, due to speech variability (e.g., different accents)~\citep{benzeghiba2007automatic} or when there is a large number of application-specific vocabulary/nouns that were not present in the training data (as is often the case with voice queries). This has motivated research into ASR error detection and correction~\citep{errattahi2018automatic}, which we classify into three main groups: study of the circumstances that lead to ASR errors, ASR error detection and ASR error correction.

The first line of work in this direction studies where ASR systems tend to produce errors in different applications. For example, early work by~\cite{voll2006methodology} identified a collection of heuristics that significantly improve ASR errors in the domain of radiology. \cite{goldwater2010words} identified {\em doubly confusable pairs} (acoustically similar words that also have similar language model probabilities), as the main source of ASR errors. 

Concerning work on ASR error detection, \cite{hirschberg2004prosodic} showed that using prosodic features can predict ASR errors more accurately than using acoustic features. Later work by \cite{litman2006characterizing} showed that it is possible to train machine learning models to predict ASR errors with better than random accuracy from features available to the ASR system 
and later work showed this can be improved with additional features~\citep{chen2013asr} 
Language models have also been used for this purpose~\citep{tam2014asr}. An overview of these techniques can be found in the work of \cite{errattahi2018automatic}.

Most related to the work presented in this paper is the area of ASR error correction. Early work \cite{lin2000error} proposed using domain-specific pre-parsed ``exemplar sentences'', which might be enough for closed domains with a small number of possible intents. Another approach is using a statistical machine translation model trained on raw/corrected ASR output pairs to correct future errors~\citep{cucu2013statistical}. Recent work uses language models to correct the output of ASR systems 
For instance, \cite{guo2019spelling} used LSTM-based models for this purpose, and \cite{hrinchuk2020correction} used BERT~\citep{devlin2018bert}. While this is an interesting approach that achieves good performance in datasets similar to those on which language models are trained, notice that common uses of ASR systems involve users making internet, or online service search queries. These queries have very different statistical properties than standard language corpora, leading to increased error rate. 
Training these language models on corpora with the expected distribution can help significantly reduce error. For example \cite{mani2020asr} train a machine translation model on doctor patient conversations, which is applied to the output of two commercial ASR systems, achieving a 7\% word error rate reduction in this domain.

The most related line of work to {\em Mondegreen} is the work by \citeauthor{ponnusamy2020feedback}~\cite{ponnusamy2020feedback}, in the context of Amazon's Alexa. \citeauthor{ponnusamy2020feedback} collect a dataset consisting of sequences of user interactions with Alexa. 
A Markov Model is built by first translating each user query to its corresponding intent using Alexa's NLU system, and then query correction works by identifying the interpretation of the current query that maximizes the probability of reaching a successful terminal state in the Markov Chain. 
In contrast, in Mondegreen, we do not translate utterances to intents, and work directly on phonetic transcriptions of the audio voice queries.

\section{Voice Query Correction}\label{sec:task}

Once deployed,
user queries might come from a fairly large set of different ASR sources, both first and third party. Given we cannot control what third party ASR systems produce, just improving the first party ASR system is not enough. Thus, Mondegreen aims at solving the problem of ``ASR output error correction''. Which is defined as follows:
\begin{itemize}
    \item Input: a textual query (the raw output of the ASR system, plus some metadata, including which ASR system produced the transcription and timestamp).
    \item Output: an error-corrected textual query. If the system believes the query was correctly recognized by speech recognition, it should just output the unmodified input text query, but if the system believes there was some speech recognition error, a corrected query should be returned. 
\end{itemize}

\begin{table}[tb]\centering 
\begin{tabular}{l|l|l} 
{\bf Input} & {\bf Output} & {\bf Notes} \\ \hline 
``gaming chair''    & ``gaming chair''  & Not rewritten \\
``rocks and''       & ``roxanne''       & Rewritten\\
``look out music''  & ``work out music'' & Rewritten\\ 
``wacom down''      & ``walk them down'' & Rewritten\\
``how stores''      & ``house tours'' & Rewritten\\
\end{tabular}
\caption{Sample input/output pairs from our training data.}
\label{tbl:examples}
\end{table}

We note that improvements could be achieved by taking into account the user query history, but we leave that for future work. To illustrate the problem, Table \ref{tbl:examples} shows input-output examples from our training data. Notice that the first example (``gaming chair'') did not require any error correction, and thus the system should output the unmodified input query. However, the other pairs require error correction. The ``rocks and'' $\to$ ``roxanne'' example illustrates a particularity of the language distribution commonly used in queries, which contains a large proportion of proper nouns, when compared to standard text corpora used to train language models.

ASR errors contribute significantly to query abandonment. Moreover, as mentioned earlier, although ASR has improved significantly over the past few years, the language distribution used to train existing ASR systems could be very different from the distribution observed in user voice queries. The remainder of this section briefly describes the dataset used to train Mondegreen, how is it different from existing public domain ASR datasets, and then a statistical comparison of the language distributions in these two types of data to show the significant differences.

\subsection{Mondegreen Training Data}

We collected training data from transcribed voice search queries issued to one of the largest search systems in Google by users. 
The version for which we report results in this paper was trained with a dataset of 30M queries. Queries came from devices whose settings are set to the English/US region, which were the largest fraction of the traffic at the time. Our training data does not contain any audio signals. For each query, we record the text transcription generated by the ASR systems, and annotate them with timestamps, ID of the ASR system that transcribed the query, as well as information about which of the associated search results was clicked if any.

This data is significantly different from existing datasets to train ASR systems. For example, two common public datasets are {\em LibriSpeech}~\citep{panayotov2015librispeech} and {\em Common Voice}\footnote{\url{https://commonvoice.mozilla.org/}}. LibriSpeech was used for example in the work of ASR error correction of \cite{hrinchuk2020correction} and \cite{guo2019spelling}. However, these datasets contain data from audiobooks (in the case of LibriSpeech), or standard language in the case of Common Voice. The main problem with these dataset is that they do not reflect the language distribution used in user voice queries, which, for example, usually contain large number of proper nouns, as we show below.

\subsection{Language Distribution}

\begin{figure}[t!]
	\includegraphics[width=\columnwidth]{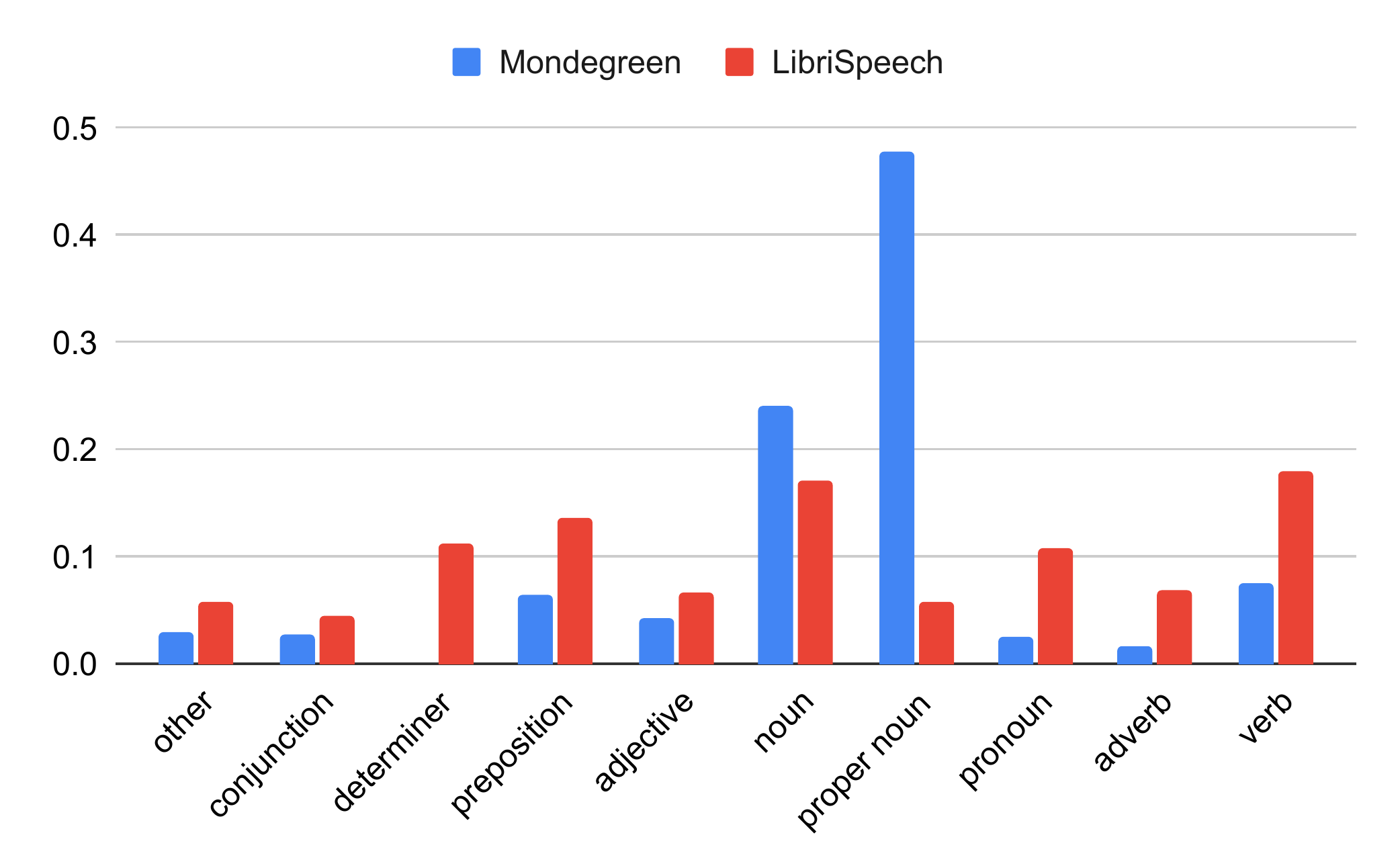}
	\centering
	\caption{Comparison of the part-of-speech (POS) distribution on a standard ASR text corpus (LibriSpeech), compared to that in user voice queries (Mondegreen).}
	\label{fig:datasets}
\end{figure}



In order to obtain a clear picture of how different is the language distribution found in voice user queries from that in datasets typically used to train ASR systems, we computed the part-of-speech (POS) tag distribution and the average sentence length. 
We compared two datasets: LibriSpeech (described above), and the dataset used to train Mondegreen. In order to perform these analyses, we used the word tokenizer and POS tagger in the Google Cloud Natural Language API. 
Given the large set of different POS tags recognized by this system, we grouped them into 10 large groups (for example, all the different pronoun types are shown together just as ``pronoun'').

Figure \ref{fig:datasets} shows the POS distributions for LibriSpeech and Mondegreen, highlighting how user voice queries (Mondegreen) are composed mostly of proper nouns and nouns with few other POS tags being represented, whereas standard datasets like LibriSpeech contain a much more even distribution of POS tags. Moreover, we note that many proper nouns used in queries are probably not recognized by the POS tagger as such, such as artist names like ``xxtenations'' or ``6IX9INE''.

Moreover, the average sentence length in LibriSpeech is 10.609 words, where as the average query length is only 3.890 words.


These differences are important, since third party ASR systems might be trained in datasets with a similar distribution to LibriSpeech, which might cause larger ASR errors when used to transcribe user voice queries. Again, although it is possible to train a 1P ASR system with in-domain data, we have limited control over the training data of 3P ASR systems. As such, Mondegreen is still needed to deal with queries coming from 3P ASR systems.

\section{Mondegreen}\label{sec:mondegreen}

Fundamentally, Mondegreen is a statistical error correction system for voice queries trained directly from sequences of user queries, described below. Mondegreen is used as a post-processing solution to correct ASR errors.
Given the dataset described above, containing a set of queries $Q$, we construct a training set $D$ consisting of {\em user corrections to abandoned voice queries} as follows. First, we divide queries into two main groups:
\begin{itemize}
    \item {\em Successful Queries}: a successful query is one for which the user eventually clicked in one or more of the search results resulting from the query.
    \item {\em Abandoned Queries}: queries that were not successful.
\end{itemize}

Each training instance in $D$ is a pair of queries $(q_1, q_2)$, where $q_1$ was {\em abandoned} and $q_2$ (the correction) was {\em successful}. To construct these pairs, we selected pairs of abandoned-successful queries that a user issued with less than $t$ seconds of difference (e.g. $t=60$). Notice that this might include pairs where $q_2$ was not an actual correction to $q_1$, we explain how we filter those below.

To train Mondegreen, we compute the following tables:
\begin{itemize}
    \item $\mathit{count}(q_2|q_1)$: the number of times in our dataset we have seen $q_2$ being proposed as a correction for $q_1$,
    \item $\mathit{count}(q)$: the number of times $q$ appears in $Q$, and
    \item $\mathit{abandonment\mhyphen rate}(q)$: the proportion of times that query $q$ was abandoned in $Q$.
\end{itemize} 
%
%
These counts result from ``exact query matching'' after normalizing the queries (lower-casing, and removing unnecessary spaces).

Then, given a new user query $q$, Mondegreen determines if $q$ needs to be corrected and how, by first retrieving a set of candidate corrections $C$ satisfying certain criteria, and then returning the most likely one. The candidate corrections are defined as follows:

\[
C = \left\{
q' \in Q \left|
\begin{array}{l}
\mathit{phonetic\mhyphen distance}(q', q) \leq \tau \,\,\, \wedge \\
1 - \mathit{count}(q'|q) / \mathit{count}(q) < \mathit{abandonment\mhyphen rate}(q) \,\,\, \wedge \\
\mathit{count}(q'|q)/\mathit{count}(q) > \beta
\end{array}
\right.
\right\}
\]
In other words, $C$ is constructed by:
\begin{itemize}
    \setlength\itemsep{0em}
    \item First selecting those queries that are phonetically similar to $q$ (distance smaller than a constant $\tau$). To calculate phonetic distance, queries are converted to sequences of phonemes, and compared using an edit-distance.
    \item Only re-writes $q'$ that are less likely to be abandoned than the original query are kept ($1 - \mathit{count}(q'|q) / \mathit{count}(q) < \mathit{abandonment\mhyphen rate}(q)$).
    \item And finally, uncommon corrections (those that occur less than $\beta \in [0,1]$ proportion of the times) are filtered out (e.g., 0.1 or 0.2).
\end{itemize}

If $C$ is empty, Mondegreen will not propose any correction. Otherwise,  $q^* = \mathit{argmax}_{q' \in C} \mathit{count}(q'|q)$ is returned as the correction, which represents the most common correction for $q$ from the set of candidate corrections.
Moreover, in the experiments presented below, we only trained on those query pairs from $D$ where the abandonment rate of $q_1$ is higher than a threshold $\alpha \in [0,1]$. 






\subsection{Engineering Considerations}

\begin{figure}[t!]
	\includegraphics[width=\columnwidth]{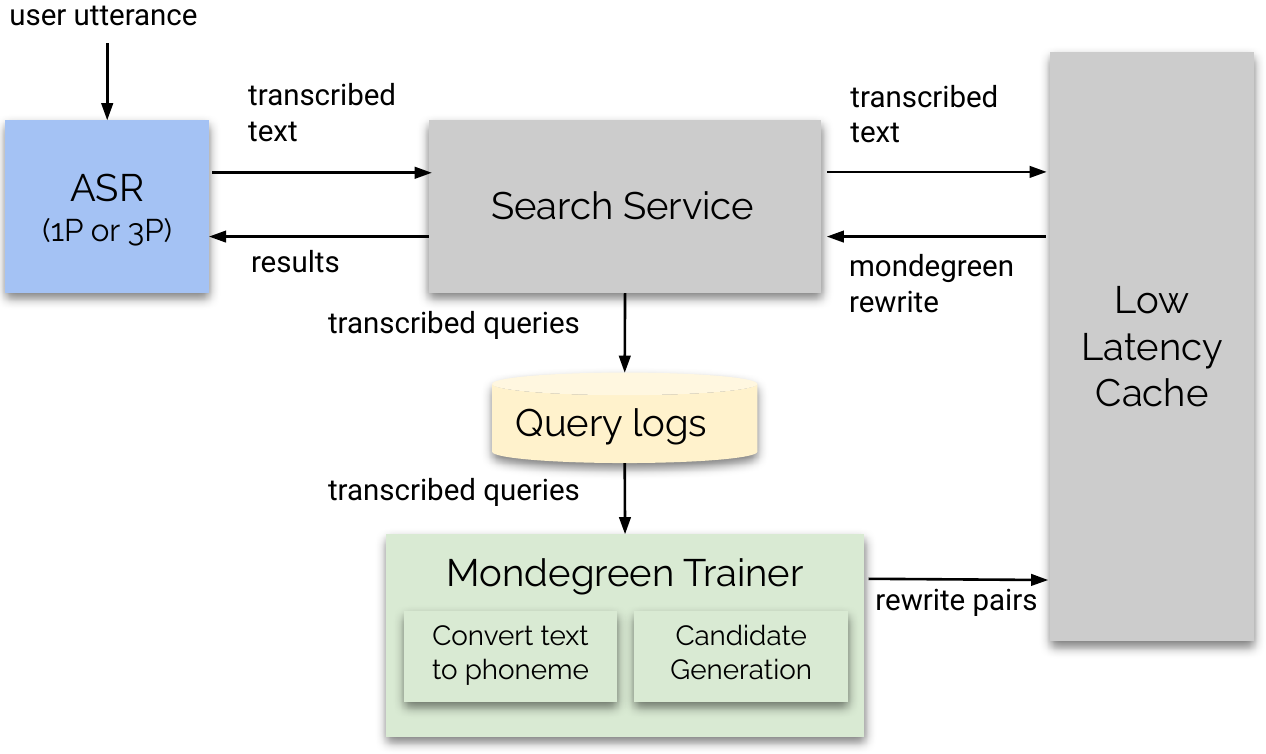}
	\centering
	\caption{High-level overview of the Mondegreen implementation, showing the offline model trainer, which stores the query rewrite pairs to a low latency cache, which is used at run-time to propose query corrections.}
	\label{fig:diagram2}
\end{figure}

This section discusses some of the lessons learned from the engineering effort required to design Mondegreen, as well as from several iterative deployments. 

The first issue with earlier versions of Mondegreen was narrowing down the right statistics to track, as soon as refinement rate and abandonment rate were made a centerpiece of the model the quality improved significantly.

We had twin goals of high precision with high coverage (trigger rate). However, query rewriting is very intrusive, so we had to ensure that rewrites were very high precision, this came at the expense of coverage.
We solved the coverage problem by focusing on showing suggestions (rather than direct rewriting) for a larger number of voice queries that were likely to be abandoned. Since suggestions are less intrusive, we could show them even when the model was not very confident (however, we are not evaluating this component in this paper).

Moreover, although our goal was very high precision, the model could not be correct 100\% of the times. So, in the real deployed version, we add an ``escape hatch'' that allows the user to go back to the original transcription.

Concerning the challenges, the main challenge was the lack of access to ground truth audio, and the difficulty of obtaining human labeled data to identify queries that are phonetically similar. This made internal iterative evaluation hard.

Concerning the technical system design of Mondegreen, as Figure \ref{fig:diagram2} shows, the offline Mondegreen trainer stores all the query rewrites as key-value pairs in a low latency distributed cache that is geolocated with the service in which the model was deployed. This ensures queries can be rewritten without any significant latency increase perceived by the user.

\section{Comparison with a Neural Model}\label{sec:experiments-neural}

\begin{table}[tb]
\begin{tabular}{ll|r|rl} 
& {\bf Model} & {\bf Test Set} & {\bf BLEU} & \\  \cline{2-4}
& No correction   & complete set   &   0.65 \\
& NMT         & complete set    & 0.68 & \\
& NMT         & triggered set  & 0.59 & \\
& Mondegreen & triggered set  & 0.79 & \\
\end{tabular}	
\caption{Comparison of corrections issued by Mondegreen with those issued by a Transformer model.}
\label{tbl:nmt-comparison}
\end{table}

We first performed an offline evaluation, comparing Mondegreen to a neural network model. The neural model was trained on three months of data. From these three months of data, we generated a training set where 50\% of the data consisted of abandoned-successful query pairs, and 50\% of the data consisted of queries that were successful to begin with, and thus the system has to learn not to propose any correction. 10\% of this data was held out as a test set (\emph{complete set}), and we also separate a subset of the \emph{complete set} of which Mondegreen actually proposed corrections ({\em triggered set}). We evaluate performance on both and report the BLEU score~\cite{papineni2002bleu} of the proposed correction with respect to the corresponding successful query. We compared against a Transformer-based~\cite{vaswani2017attention} neural machine translation (NMT) model trained to predict a successful query given an abandoned query.

Table \ref{tbl:nmt-comparison} shows our empirical results. Our baseline results (not performing any correction on the queries) results in a 0.65 BLEU score, which the NMT approach can improve to 0.68. Meanwhile, Mondegreen triggers only on a small subset of queries, but when it triggers, Mondegreen's corrections are very high quality: BLEU score 0.79. Notice that the NMT model struggled in those queries (0.59 BLEU) highlighting that, while being conceptually and computationally simple, Mondegreen's predictions are a great complement to those produced by the NMT model.

\section{Live Experiment}\label{sec:experiments}

We evaluated Mondegreen on one of the largest voice search systems from Google in an A/B live experiment. This section first presents the experiment setup, followed by the results.

\subsection{Setup}

The control group is the production voice search system with highly optimized 1P and 3P ASR systems. The experiment group is the production voice search system with Mondegreen enabled to correct transcribed queries.
The results presented below are evaluated over 2M queries. Moreover, we would like to highlight that the results reported here used commercial grade 3P ASR systems that have been perfected over many years by many proprietary providers. Thus, all the improvements Mondegreen shows are over these highly optimized ASR systems that are used widely.
In order to evaluate the performance of Mondegreen, we report the following metrics. 
\begin{itemize}
    \item {\em Trigger rate}: percentage of voice queries that Mondegreen replaces with a correction.
    \item {\em Click Through Rate (CTR)}: Number of search results clicked per query times 100.
    \item {\em User Interaction}: The level of interactions between users and the search results (number of clicks resulting in extended user interaction).
    \item {\em Abandoned voice queries}: percentage of all voice queries that were abandoned (have zero clicks and no refinement).
    \item {\em Voice queries with refinement}: percentage of queries that have a follow up within a phonetic similarity threshold.
    \end{itemize}

Percentage differences and confidence intervals were calculated using the {\em Pre-Post} method~\cite{soriano2017percent}, specially designed for large online experiments.
In order to highlight the metrics where Mondegreen achieved a statistically significant improvement, we calculated the 95\% confidence interval for all metrics, and we will highlight those results where the metrics for the Control group is outside the Mondegreen confidence interval.

Moreover, in order to gain further insights into these metrics, and how do they affect different user groups (depending on their region, or which device they use to connect) we will also present breakdowns by whether queries came from a first party ASR (1P), or from a third party ASR (3P), as well as breakdown by language region (US, UK, India, which are three of the largest regions where the version of Mondegreen presented in this paper was deployed). 

\subsection{Results}


\begin{table}[tb]\centering 


\resizebox{\columnwidth}{!}{%
\begin{tabular}{l|ccc} 
& {\bf \% A/B Test Diff.} & {\bf Notes} \\ \hline
CTR                 & 4.73\% (-0.61\%, +10.17\%)   & \\
User Interaction    & {\bf 9.49\% (+2.77\%, +16.47\%)} & \\
Abandoned queries        & -1.90\% (-6.67\%, +3.05\%) & lower is better \\
Queries w. refinement  & {\bf -7.09\% (-11.09\%, -2.98\%)} & lower is better 
\end{tabular}
}
\caption{Overall results on queries where Mondegreen proposed a correction (a 1.76\% of the total traffic). Numbers shown in bold represent statistically significant differences. Confidence intervals are shown in between parentheses.}

\label{tbl:experiment1}
\end{table}

Table \ref{tbl:experiment1} shows the metrics for the comparing the control group to the Mondegreen group. Mondegreen proposes corrections to 1.76\% of the total traffic, and when it proposes corrections, it improves the quality of the search results significantly. Moreover, notice that 1.76\% is not as low a percentage as it might seem since, first, a large proportion of queries are correctly interpreted by the ASR system to begin with (and hence, we don't want to propose corrections to those) and second, 1.76\% of the total traffic of a large search system is a very large number of queries in absolute terms for which we are improving the search results.

Table \ref{tbl:experiment1} shows that in the set of queries for which Mondegreen triggers, the User Interaction increased by 9.49\%. Moreover, the percentage of queries for which the users issued refinements reduced by 7.09\%. Both these differences are statistically significant. 
Notice that reducing the number of refinements required by users is a very desirable effect. Even if not all refinements are bad (e.g., when the user is searching interactively moving forward with her interaction), many refinements result from misunderstood voice queries, which are the refinements Mondegreen is targeting to reduce.

Additionally, the set of queries for which Mondegreen triggers are more likely to be abandoned than the average query. Even thought not shown in the table, the evidence for this claim is that the average CTR for those queries where Mondegreen would have triggered (in the control group) is 21.20\% lower than the average CTR for those queries where Mondegreen would not have triggered. This means that users click on average 21\% less search results for these queries, confirming that the triggering conditions of Mondegreen successfully identify a subset of queries that are indeed more likely to be abandoned.

In conclusion, Mondegreen triggers for a small percentage of queries, but when it triggers, it results in significant improvements.

\subsection{Breakdown by ASRs}

\begin{table}[tb]\centering 






\resizebox{\columnwidth}{!}{%
\begin{tabular}{l|ccc} 
& {\bf \% A/B Test Diff.} & {\bf Notes} \\ \hline \hline
& \multicolumn{2}{c}{1P ASR (about 25\% of the test data)} \\ \hline
CTR                 & {\bf -11.70\% (-21.84\%, -0.18\%)}   & \\
User Interaction    & -8.71\% (-20.16\%, +3.66\%) & \\
Abandoned queries        & 15.02\% (-3.58\%, +35.97\%) & lower is better \\
Queries with refinement  & -3.97\% (-16.40\%, +9.83\%) & lower is better \\ \hline 
\hline
& \multicolumn{2}{c}{3P ASR (about 75\% of the test data)} \\ \hline
CTR                 & {\bf 7.31\% (+1.67\%, +13.19\%)}   & \\
User Interaction    & {\bf 13.09\% (+5.90\%, +20.67\%)} & \\
Abandoned queries        & -3.81\% (-9.03\%, +1.63\%) & lower is better \\
Queries with refinement  & {\bf -7.37\% (-11.73\%, -2.84\%)} & lower is better \\ \end{tabular}
}
\caption{Overall results on queries impacted by Mondegreen, but broken down by first or third party ASR.}
\label{tbl:experiment-asr}
\end{table}

Table \ref{tbl:experiment-asr} shows experimental results comparing the performance of Mondegreen on different ASR systems. We divided the results by first party ASR (1P), versus third party ASRs (3P). The first thing we can see is that Mondegreen significantly helps in 3P queries (which were the majority, about 75\% of our data, during the period of time in our study). For example, in 3P queries, CTR goes up 7.31\% for those queries where Mondegreen triggered, user interaction went up 13.09\%, and the number of abandoned queries and queries with refinements went down.

The Mondegreen results on 1P queries show slight deterioration. This is due to the imbalance in training data which skews towards 3P queries. The same can be addressed by including more 1P queries in the training or increasing the weight of 1P queries which is left as future work. 
However, even if the table does not show this, overall the quality of the search results provided by the first party ASR is significantly higher to begin with, and hence the absolute values for CTR and User interaction are much higher to begin with than those coming from third party ASRs. Thus, as Table \ref{tbl:experiment1} shows, the overall effect of Mondegreen, even in the version reported in this paper, is a significant increase in search results (higher CTR, less abandoned queries, and less queries with refinements).

In summary, we see that Mondegreen seems to help the most for third party ASRs where users tend to click on fewer search results, abandon more queries and need more query refinements.

\subsection{Breakdown by Locale}



\begin{table}[tb]\centering 

\resizebox{\columnwidth}{!}{%
\begin{tabular}{l|ccc} 
& {\bf \% A/B Test Diff.} & {\bf Notes} \\ \hline \hline
& \multicolumn{2}{c}{US (about 50\% of the data)} \\ \hline
CTR                 & 4.48\% (-1.75\%, +11.04\%)   & \\
User Interaction    & {\bf 8.41\% (+1.15\%, +16.11\%)} & \\
Abandoned queries        & -1.16\% (-6.48\%, +4.40\%) & lower is better \\
Queries with refinement  & {\bf -7.75\% (-12.37\%, +2.93\%)} & lower is better \\ \hline 
\hline

& \multicolumn{2}{c}{India (about 20\% of the data)} \\ \hline
CTR                 & 8.24\% (-3.80\%, +22.58\%)   & \\
User Interaction    & 10.07\% (-7.06\%, +29.29\%) & \\
Abandoned queries        & -7.75\% (-18.59\%, +4.71\%) & lower is better \\
Queries with refinement  & -9.35\% (-23.43\%, +6.52\%) & lower is better \\ \hline \hline

& \multicolumn{2}{c}{UK (about 10\% of the data)} \\ \hline
CTR                 & {\bf 20.01\% (+2.50\%, +39.95\%)}   & \\
User Interaction    & 17.09\% (-3.22\%, +42.58\%) & \\
Abandoned queries        & -1.83\% (-16.22\%, +15.46\%) & lower is better \\
Queries with refinement  & -10.73\% (-22.45\%, +32.20\%) & lower is better \\ \end{tabular}
}
\caption{Overall results on queries impacted by Mondegreen, but broken down by locale. We include only three of the largest locales where Mondegreen was deployed.} 

\label{tbl:experiment-locale}
\end{table}

Finally, Table~\ref{tbl:experiment-locale} shows results for different locales. Notice that our model was trained on data only from English queries with US locale, but it was deployed in other English locales (India, UK, etc.). So, it was not obvious that a model trained only on US data would help in other locales. We report the three English locales with the largest amount of traffic: US, India, and UK. 

We see that Mondegreen seems to help in all locales, as can be seen by increased CTR and User Interaction, although only the increase in User Interaction for US, CTR for UK, and queries with refinement for US are statistically significant. As a matter of fact, it is in the UK where the larger gains are observed, with CTR going up 20.01\% for those queries where Mondegreen triggered.


The conclusion is that different locales of English share a lot of commonalities, and hence, a model trained on the US locale seems to still provide benefit for non-US locales.

\section{Conclusions}\label{sec:conclusionss}

This paper presented Mondegreen, a system to correct automatic speech recognition (ASR) errors for user voice queries. Mondegreen is a statistical approach based on calculating the probability that a query correction is more likely to be successful than the original query issued by the user. 

We showed that the language distribution exhibited by user voice queries is significantly different than the distribution on standard text corpora used to train ASR systems, which can explain the increased number of ASR errors seen in practice, leading to query abandonment. Additionally, we evaluated Mondegreen against a Transformer model, as well as in live-experiments with real users in a major Google search product. The results show that despite being conceptually simple (and computationally efficient), Mondegreen can complement highly optimized production ASR systems and achieve significant gains. Our results show that queries coming from third-party ASR APIs benefit the most, and also that even if the version of Mondegreen reported in this paper was only trained on US data, gains can be obtained in other locales (such as UK).



\bibliographystyle{ACM-Reference-Format}





\end{document}